# Thermoelectric power factor of Bi-Sb-Te and Bi-Te-Se alloys and doping strategy: First-principles study

Byungki Ryu (류병기),[*] Jaywan Chung (정재환), Bong-Seo Kim (김봉서), and Su-Dong Park (박수동)

[1]Thermoelectric Conversion Research Center, Korea Electrotechnology Research Institute (KERI), Changwon-si, Gyoengsangnam-do, 51543, South Korea

Eun-Ae Choi (최은애)

[2]Korea Institute of Materials Science, Changwon-si, Gyoengsangnam-do, 51508, South Korea

* Corresponding author. E-mail address: byungkiryu@keri.re.kr (B. Ryu)

Abstract: By performing first principles calculations combined with Boltzmann transport equations, we calculate the thermoelectric power factor (PF) of $(Bi_{1-x}Sb_x)_2Te_3$ (BST, $0 \leq x \leq 1$) and $Bi_2(Te_{1-y}Se_y)_3$ (BTSe, $0 \leq y \leq 1$) ternary alloys as a function of alloy composition ratio, carrier concentration, and temperature. The structure relaxation and the mixing entropy can stabilize the ternary solid solution phases. For p-type BST, the thermoelectric performances of ternaries are comparable to the $Bi_2Te_3$ and the maximum PF is found at the hole concentration near $4\times10^{19}$ cm$^{-3}$. For n-type BTSe, the thermoelectric performances are composition and configuration dependent and the optimal carrier concentration is similar or higher than that for BST. When y is less than 1/3, the PFs of BTSe are comparable to $Bi_2Te_3$. However, as y approaches 1, the thermoelectric performance reduces. We also find that the thermoelectric performance of BST is superior to that of BTSe due to the longer electron relaxation time for BST and the small valley band degeneracy of the $Bi_2Se_3$ conduction band. The electron transport anisotropy is higher for BTSe (~4.8) compared to BST (~2.3), due to the poor electric conduction along out-of-plane direction in BTSe. We also investigate the effect of temperature on the PFs. For p-type BST, the band gap effect on PF is relatively small for BST and PFs at optimal carrier concentration are decreasing with increasing temperature. For n-type BTSe, the PFs at optimal doping range are maintained until temperature is less than 400 or 500K. The optimal doping concentration for p-type BST is about $4\times10^{19}$ cm$^{-3}$, which is achievable by Sb alloying. The optimal doping concentration for n-type BTSe is about $6\times10^{19}$ cm$^{-3}$ or higher, which needs additional extrinsic dopant in addition to Se alloying. The point defect formation energy calculations reveal that Cl, Br, and I impurities are potential candidates for n-type carrier source, while F as well as Au is the compensating defect.

## 1. Introduction

Thermoelectricity refers to the direct energy conversion between heat and electricity, and its technology can be applicable to the thermometer, power generator, and refrigerator [1]. For the thermoelectric application, the high performance and efficiency of energy conversion is desirable. The efficiency of thermoelectric conversion is determined by the dimensionless parameter, the thermoelectric material figure of merit ZT, defined as ZT = $(\sigma\alpha^2/\kappa)$T, where $\alpha$, $\sigma$, $\kappa$, and T are Seebeck coefficient, conductivity, thermal conductivity, and absolute temperature [1,2].

Alloying is one of the best routes to obtain high ZT thermoelectric materials with several reasons. Material alloying can reduce the thermal conductivity in phonon contribution ($\kappa_{ph}$) by enhancing the phonon-scattering event while phonon transports [3, 4]. Moreover, in alloys, nanoprecipitation can act as a phonon scatterer suppressing the phonon transport with mean free path length in order of 10 to 100 nm [5-9]. Hierachical disorders have been succeeded to obtain very low thermal conductivity [11]. It also optimizes the electrical properties by enhancing density of states effective mass thereby enhancing the Seebeck coefficient through the formation of resonant level [12] or through the band convergence [13,14].

$Bi_2Te_3$ is one of the best thermoelectric materials performing at the temperature range from 200 to 500 K [1,2]. It exhibits high band degeneracies with narrow gap semiconducting nature in addition to strong spin-orbit-interacting effect [15-21]. As a result, it carries high power factor (PF) of about 3 to 5 mW/m/K$^2$ near the room temperature under p-type condition [22-24]. For $Bi_2Te_3$, the alloying is the important process to optimize the thermoelectric properties. By alloying with $Sb_2Te_3$ and $Bi_2Se_3$, we can tune the carrier concentration and the position of Fermi level, and finally obtain the p-type and n-type thermoelectric materials [1,2,22-25]. Of course, the $\kappa_{ph}$ is also reduced with alloying [3].

In previous study, we reveal that the alloying also affects the band structure of $Bi_2Te_3$: The band characteristics, such as band gap, band edge positions, and band degeneracies are changed by alloying [19,20]. $Bi_2Te_3$, $Sb_2Te_3$, and $Bi_2Se_3$ have similar valence band structures, as compared to the conduction band structures. As a result, the change of band structure is significant for n-type region. Meanwhile, the change of low energy valence band characteristics is rather small. In addition, there are some studies about the effect of strain or band gap after alloying [26-29]. Alloying with $Sb_2Te_3$ found to be important due to the increase of the intrinsic carrier and the decrease of the bipolar thermal conductivity [29].

Like this, there have been lots of theoretical studies as well as experimental studies on the thermoelectric properties of $Bi_2Te_3$ *"binary"* alloys. However, still there is lack of theoretical study on the thermoelectric properties of $Bi_2Te_3$-related *"ternary"* alloys. Even though the alloying is very important to obtain high performance thermoelectric materials, in our knowledge, only certain stoichiometry with ordered structure is investigated theoretically [30,31]. There is no systematic theoretical study of thermoelectricity for alloy composition ratio in $Bi_2Te_3$. The effect of alloying on the PF is also not clear yet.

In this work, to investigate the thermoelectric properties of $(Bi_{1-x}Sb_x)_2Te_3$ (BST) and $Bi_2(Te_{1-y}Se_y)_3$ (BTSe) alloys, we perform the first principles density functional theory calculations and the Boltzmann transport calculations. We generate the model structure for BST and BTSe alloys in solid solution. Then, we calculate the thermoelectric PFs for BST and BTSe alloys as a function of composition ratio, carrier concentration, and temperature. To reach the maximum power factor, we search the optimal dopant for shallow donors. Halgoen atoms such as Cl, Br, and I are found to be efficient donors, while F as well as Au is not.

## 2. Computational Approach

We perform the first principles density functional theory (DFT) calculations [32,33] to calculate the electronic band structures of BST and BTSe. We use the projector-augmented-wave (PAW) pseudopotentials [34], generalized-gradient-approximation (GGA) exchange-energy functional [35], planewave basis, which are implemented in VASP code [36,37]. We include the spin-orbit-interaction, which is important to describe the low energy band structures of $Bi_2Te_3$-related heavy materials. For binary compounds, we use the 5-atom rhombohedral primitive cell with gamma centered 36×36×36 k-point mesh containing 4237 irreducible kpoints. For ternary compounds, we use the 20-atom (2×2×1) rhombohedral supercell with gamma centered 18×18×18 k-point mesh. The number of irreducible kpoints of ternary supercells is varying between 3089 and 5836, depending on the supercell symmetries. For alloy atomic structures, we consider the $Bi_{8-m}Sb_mTe_{12}$ and $Bi_8Te_{12-n}Se_n$ supercells where m = 0, 1 , … , 8 and n = 0, 1 , … , 12, respectively. The positions of Sb atoms are chosen randomly. The lattice parameters and the internal coordinates of ternary systems are linearly interpolated from the values from binary systems, as we did in previous work [19].

The thermoelectric properties ($\alpha$, $\sigma$, $\kappa_{el}$, power factor $\alpha^2\sigma$) of BST and BTSe are calculated as a function of composition ratio x and y, carrier concentration ($n$) and temperature by using Boltzmann transport equation within a constant relaxation time approximation and a rigid band approximation, which is implemented in BoltzTraP code [38,39]. Note that the band gap is very important due to the bipolar transport at relatively high temperature [40,41]. To overcome the band gap underestimation in DFT-PBE calculations, here we use the experimental or GW corrected band gaps of 0.16, 0.23, 0.30 eV for $Bi_2Te_3$ [42], $Sb_2Te_3$ [43], and $Bi_2Se_3$ [44-46] binaries, respectively. For ternaries, linearly interpolated band gaps are used. To provide the reliable thermoelectric properties compared to experiment, we estimate the electron relaxation time ($\tau$) by fitting the computational thermoelectric property pairs ($\alpha$, $\sigma$) of $Bi_2Te_3$ at 300 K to the experimental properties of BST [29,47,48,49,50] and BTSe [51,52,53] and use the $\tau = \tau_O \times (300K/T)$ ($\tau_O = 1.8 \times 10^{-14}$ sec for hole and $\tau_O = 1.8 \times 10^{-14}$ sec for electron).

To compare the stability of defect configuration, we calculate the formation energy [ref]. The defect formation energy of M point defect is calculated as $E_{FORM}[Bi_2Te_3:M] = E_{tot}[Bi_{96-x}Te_{144-y}M] - 48\ E_{tot}[Bi_2Te_3] + \{\ x\ \mu[Bi] + y\ \mu[Te] - E[M]\}$ [54,55], where $E_{tot}[Bi_{96-x}Te_{144-y}M]$, $E_{tot}[Bi_2Te_3]$, $\mu[Bi]$, and $\mu[Te]$ are total energy of defective supercell, total energy of $Bi_2Te_3$, atomic chemical potential of Bi, and atomic chemical potential of Te. Here atomic chemical potentials are constrained as $2\ \mu[Bi] + 3\ \mu[Te] = E_{tot}[Bi_2Te_3]$, $\mu[Bi] \le E[Bi]$ and $\mu[Te] \le E[Te]$, where $E[Bi]$ and $E[Te]$ are the reference energy of Bi and Te bulk in rhombohedral phases.

# 3. Results and discussions

The lattice structure of $Bi_2Te_3$ is a tetradymite structure with a group number of 166, having a rhombohedral primitive unitcell [19]. In the primitive cell, there are two Bi and three Te atoms. The atomic layer of Te(1)-Bi-Te(2)-Bi-Te(1) in $Bi_2Te_3$ are stacked as (ABCAB)(CABCA)(BCABC) sequence and 15-atmic layerse consist one hexagonal unit. Due to the weak bonding nature between the Te(1) layer and the adjacent Te(1) layer, there is a large distance between them and five layers form a one quintuple layer (QL) [20,28]. Each QL consists of two Bi at the internal coordinate of (**u u u**), two Te(1) at (**v v v**), and one Te(2) atoms at (0,0,0). Thus, there are four independent lattice parameters describing $Bi_2Te_3$-related materials: lattice parameter **a** and **c**, and the internal coordinates **u** and **v**. Here we use the experimental lattice parameters or the interpolated lattice parameters [19].

We first investigate the thermoelectric properties of $Bi_2Te_3$ as a reference material. **Figure 1(a)** shows the Pisarenko plot, Seebeck coefficient $\alpha$ as a function of carrier concentration (**n**) for $Bi_2Te_3$ at 300 K. Here we use the directional average of $\langle\alpha\rangle$ for polycrystalline limit, i.e. $\langle\alpha\rangle = (\alpha_{xx} + \alpha_{yy} + \alpha_{zz})/3$, where $\alpha_{ij}$ is the Seebeck coefficient tensor. We compare the thermoelectric properties with two different band gaps: one is the calculated band gap ($E_g^{CALC}$), and the other is the experimental band gap ($E_g^{EXPT}$). When we use the $E_g^{EXPT}$, band structures are recalculated by using Scissor operator with $E_g^{EXPT}$ and then Boltzmann transports are calculated. As the band gap increased from the $E_g^{CALC}$ of 0.1 eV to the $E_g^{EXPT}$ of 0.16 eV, the $\langle\alpha\rangle$ at the low carrier concentration less than $10^{19}$ cm$^{-3}$ is largely enhanced due to the less density of minority carriers. When **n** = $10^{18}$ cm$^{-3}$, the size of $\langle\alpha\rangle$ is about 300 to 400 µV/K for experimental gap and only about 100 µV/K or smaller for the calculated gap result. **Figure 1(b)** shows the temperature dependent $\langle\alpha\rangle$ with the experimental band gap. For a low doping regime, **n** < $10^{19}$ cm$^{-3}$, the $\langle\alpha\rangle$ is decreased with increasing T due to the enhanced bipolar transport effect. We can rewrite the $\alpha$ as $\alpha = (\sigma_p\alpha_p + \sigma_n\alpha_n)/(\sigma_p + \sigma_n)$, where p denotes the type of carriers. Thus, at higher temperature, the electron transport through minority carrier becomes comparable to the majority carrier ($\sigma_p/\sigma_n \sim 1$) and the size of $\alpha$ is reduced because the signs of $\alpha_p$ and $\alpha_n$ are different, resulting in the reduction of $\alpha$. For high doping regime, $n > 4 \times 10^{19}$ cm$^{-3}$, the bipolar conductivity is reduced and the $\alpha$ is roughly proportional to the T. Note that, in single parabola band model, $\alpha$ is written as $\alpha = (2/3e)(k_B\hbar)^2 m^* T (\pi/3n)^{2/3}$, where g is the band valley degeneracy, $n$ is a carrier concentration, and m* is the density of states effective mass. It is known that the high ZT Bi$_2$Te$_3$ materials [24] have an optimal carrier concentration between $10^{19}$ cm$^{-3}$ and $10^{20}$ cm$^{-3}$. Thus, the effect of band gap and bipolar transport is very critical to describe the thermoelectric properties, especially for early or smaller $10^{19}$ cm$^{-3}$. Throughout this work, we use the experimental band gap for binary or interpolated band gap for ternaries from experimental values of binaries, to overcome the band gap underestimation and to describe the high temperature effect properly.

We investigate the thermoelectric properties of binary tetradymite-Sb$_2$Te$_3$, Bi$_2$Se$_3$, and Sb$_2$Se$_3$ at 300 K, with experimental lattice parameters and experimental band gaps. Although the stable geometry of Sb$_2$Se$_3$ is an orthorhombic, we consider the tetradymite structure to compare the effect of chemical stoichiometry. Figure 2(a) and (b) show the $\alpha$ as a function of $n$ for binary compounds under p and n-type conditions. For $n$ between $1\times10^{19}$ and $1\times10^{20}$ cm$^{-3}$, where $n$ is within conventional doping limit for Bi$_2$Te$_3$, the p-type performance of binary is superior to n-type performance, implying the difficulty to develop high performance n-type thermoelectric materials using Bi$_2$Te$_3$-based materials. Under p-type condition, the $\alpha$ of Bi$_2$Te$_3$ is smaller than that of Bi$_2$Se$_3$, similar to other report [45]. Under n-type condition, the best thermoelectric performance is also found in Bi$_2$Te$_3$. And the n-type performance of Sb$_2$Te$_3$ is superior to that of Bi$_2$Se$_3$. Unfortunately, however in experimentally, alloying with Sb$_2$Te$_3$ leads to p-type conduction, while alloying with Bi$_2$Se$_3$ leads to n-type conduction.

Next, we investigate the energetics of ternary alloys, BST and BTSe. We calculate the free energies of ternary alloys with respect to the binary phases, as shown in the Figure 3. We calculate the free energy G, defined as $G = H_{mix} - TS_{mix}$, where $H_{mix}$ and $S_{conf}$ are the mixing enthalpy and mixing entropy, respectively. At 0 K, the free energy corresponds to the mixing enthalpy, which is defined as $H_{mix}[AB] = E_{mix}[AB] + pV \doteq E_{tot}[AB] - E_{tot}[A] - E_{tot}[B]$, where $A$ is Bi$_2$Te$_3$ and $B$ is Sb$_2$Te$_3$ or Bi$_2$Se$_3$. The $S_{mix}$ of BST and BTSe is calculated as $S_{mix}/Nk_B = -x \ln x - (1-x) \ln (1-x)$, where N = 2 and 3 for (Bi$_{1-x}$Sb$_x$)$_2$Te$_3$ and Bi$_2$(Te$_{1-x}$Se$_x$)$_3$, respectively. Note that, in solids, the pV term in $H_{mix}$ is negligible, so that we only consider the $E_{mix}$ term. In Figure 3(a), $H_{mix}$ is positive for BST ternaries even after structural relaxation. However, $H_{mix}$ becomes less than 10 meV after relaxation, implying that the entropy can stabilize the solid solution of BST ternaries, as drawn by dot-dashed lines. In other words, Bi$_2$Te$_3$ and Sb$_2$Te$_3$ are miscible. As shown in Figure 3(b), for BTSe alloys, the $E_{mix}$ is also configuration dependent. Se prefers Te(2) to Te(1)-site. We thought that it is due to the large size difference between Te and Se atoms and the formation of Bi$_2$Te$_2$Se or Bi$_2$Se$_2$Te structure when Se content increases. When T is sufficiently high (> 600 K), the free energies of almost BTSe ternaries becomes lower than zero, meaning that BTSe ternaries can be miscible.

As we discussed in the previous section, we estimate the electron relaxation time of $Bi_2Te_3$-related materials by comparing the thermoelectric properties from calculations and experiments. For hole carriers for BST, we use the electronic relaxation time of $1.8 \times 10^{-14}$ sec at 300 K. For electron carriers for BTSe, we use the electron relaxation time of $1.2 \times 10^{-14}$ sec at 300K. We assume that the conductivity is proportional to $1/T$. Then we study the thermoelectric properties such as conductivity and power factor (PF).

We calculate the thermoelectric properties of ternary BST and BTSe alloys. In the Figure 4 (a) and (b), the power factor of BST along in-plane ($PF_{inp}$) and out-of-plane ($PF_{outp}$) directions are represented as a function of hole carrier concentration. Since $Bi_2Te_3$ and $Sb_2Te_3$ have similar low energy band structures [19], the function shapes of PFs are similar for lower hole carrier concentration region less than $10^{20}$ cm$^{-3}$. However, at higher hole carrier concentration region ($>10^{20}$ cm$^{-3}$), the PF shapes are different each other, due to the different band shape of secondary conduction band minimum between $Bi_2Te_3$ and $Sb_2Te_3$ at higher energy region.

In Figure 5, we summarize the maximum power factor of BST and BTSe for the **n** less than $2 \times 10^{20}$ cm$^{-3}$. Note that, in $Bi_2Te_3$-related materials, the available carrier concentration might not exceed $2 \times 10^{20}$ cm$^{-3}$. The PFs of n-type $Bi_2Te_3$ and BTSe are much smaller than that of those of p-type $Bi_2Te_3$ and BST. It is mainly due to the small electron relaxation time in n-type materials and partially due to the poor out-of-plane transport property compared to in-plane transport. The anisotropy of power factor is severe in BTSe, as compared to BST, consistent to the conductivity anisotropy reported by B.Y. Yavorsky and coworkers [56]. For BST and BTSe, the values of maximum $PF_{inp}$ are 2.3 and 4.3 times larger than those of maximum $PF_{outp}$, respectively. We also find that the power factor is slightly decreased after alloying. This effect can be understood from the results of binaries' thermoelectric properties. As discussed, the Seebeck coefficient of $Bi_2Te_3$ is superior to that of $Sb_2Te_3$ in p-type and that of $Bi_2Se_3$ in n-type. Also note that the reduction of PF is severe in BTSe ternaries. It is due to the reduced band degeneracy in BTSe alloys, as reported by us [19].

In Table 3 and 4, we summarize the optimal carrier concentration for maximum PF for BST and BTSe when **n** is less than $2 \times 10^{20}$ cm$^{-3}$. In BST, the hole carrier concentration of about 3 to $4 \times 10^{19}$ cm$^{-3}$ is the optimal. However, in BTSe, the optimal electron carrier concentration is varying from $4 \times 10^{19}$ cm$^{-3}$ to early $10^{20}$ cm$^{-3}$, depending on the configuration. We find that the positions of Se atoms are critical to the electronic structures and thereby thermoelectric transport properties. When Se is at Te(1) site, the optimal doping range is near $4 \times 10^{19}$ cm$^{-3}$. When Se is at Te(2) site, the optimal doping concentration exceeds $2 \times 10^{20}$ cm$^{-3}$ and the maximum $PF_{outp}$ is larger than those for Se at Te(1).

We also investigate the effect of temperature on the thermoelectric power factors of BST and BTSe ternaries. Considering the experimentally popular alloy stoichiometries, we choose x = 6/8 for BST [$(Bi_{0.25}Sb_{0.75})_2Te_3$] and y = 1/12 for BTSe [$Bi_2(Te_{1-y}Se_y)_3$]. In Figure 6, we represent the power factor at various carrier concentrations as a function of temperature for p-$Bi_2Te_3$ and p-$(Bi_{0.25}Sb_{0.75})_2Te_3$. For BST, maximum power factor is found near the **n** = $4\times10^{19}$ cm$^{-3}$ when we consider the temperature range. After Sb alloying, the $PF_{inp}$ at n = $4\times10^{19}$ cm$^{-3}$ is slightly decreased, while $PF_{outp}$ n = $4\times10^{19}$ cm$^{-3}$ is nearly maintained. Note that here we assume that the band gaps of $Bi_2Te_3$ and $Sb_2Te_3$ are 0.16 and 0.23 eV, respectively, and thereby we use the BST band gap of 0.2125 eV. Due to the larger band gap of $(Bi_{0.25}Sb_{0.75})_2Te_3$, the temperature dependency of PF is slightly different. For p-$Bi_2Te_3$, PF is monotonic decreasing when n=$4\times10^{19}$ cm$^{-3}$. For p-BST, PF has a maximum near 400K and then it is decreasing. This kind of temperature dependency is significant at the low carrier concentration region. We can clearly find the drop of PF for n = 1 and $2\times10^{19}$ cm$^{-3}$. It is more severe for smaller band gap $Bi_2Te_3$ as compared to BST. We would like to mention that there many sources of experimental band gaps for $Bi_2Te_3$ and $Sb_2Te_3$. However, due to the narrow gap nature of $Bi_2Te_3$, $Sb_2Te_3$ and $Bi_2Se_3$, the gap values are not sufficiently consistent each other. We thought that the band gap value can be extracted from the pattern of power factor or thermoelectric properties. As many studies reported, the power factors of $Bi_2Te_3$ and BST alloys always decrease with increasing T, implies that the real band gap of BST and $Sb_2Te_3$ might be smaller than the reported value. Or we expect that we can extract the exponent for temperature dependency in electron relaxation time.

In Figure 7, we represent the PFs at various carrier concentrations as a function of temperature for n-$Bi_2Te_3$ and n-$Bi_2(Te_{0.917}Se_{0.083})_3$. Among various configurations, we use the atomic model N1-2 where the Se atom is at the Te(2)-site, having lower mixing energy compared to the N1-1 configuration. At higher temperature, the power factor at 6 to $8\times10^{19}$ cm$^{-3}$ is larger than the power factor at $4\times10^{19}$ cm$^{-3}$. Thus, we may conclude that the optimal power factor for n-type materials are about $6\times10^{19}$ cm$^{-3}$ or higher. It means that we need higher doping concentration in n-type BT and BTSe, as compared to p-type BT and BST. Also we find that the PF of n-$Bi_2Te_3$ and n-BTSe at $4\times10^{19}$ cm$^{-3}$ are maintained at higher temperature, which is not observed in PF of p-$Bi_2Te_3$ at the same **n**. In other words, the bipolar effect on the reduction of thermoelectric properties in n-type $Bi_2Te_3$ and BTSe are relatively weaker than that in p-type $Bi_2Te_3$ and BST. We think that this kind of behavior is responsible for the high temperature positions above 450 K of maximum peak ZT in n-type materials, measured from experiments [57-59].

From the above results, we obtain the optimal carrier concentration to maximize PFs: $4\times10^{19}$ cm$^{-3}$ for p-type $Bi_2Te_3$ and BST and $6\times10^{19}$ cm$^{-3}$ or early $10^{20}$ cm$^{-3}$ for n-type $Bi_2Te_3$ and BTSe. It is well known that the carrier concentration of BST is near $4\times10^{19}$ cm$^{-3}$, thus there is no need to find p-type extrinsic dopant for Fermi level tuning. However, for n-type, the proper doping elements are needed to stabilize the optimal carrier concentration. Cu can be interstitial defects located between QLs and can be acting as n-type dopant [57]. Others also reported that Cu, Ag, and Au atoms are found to be acceptor or isovalent when it substitutes Bi [22,44,60]. Like this, Cu, Ag, and Au dopants can be either acceptor or donor, meaning that they are self-compensating defects. Thus, we need to find other candidate for stable n-type dopant other than Cu, Ag, or Au.

By comparing the defect formation energy ($E_{FORM}$) of halogen atoms (M= F, Cl, Br, I), we find their stable defect geometries in $Bi_2Te_3$. We consider various point impurities with various defect configurations: substitutional defects at Bi, Te(1), Te(2) sites ($M_{Bi}$, $M_{Te(1)}$, $M_{Te(2)}$) and interstitial defect ($M_{INT}$) located between QLs. For the interstitial defects, we consider the various sites including octahedral site between QLs, tetrahedral site between QLs, and the trigonal site among Te(1) in single QL. Here, to reduce the computational cost, we use the $Bi_2Te_3$ as a main matrix. Since $Sb_2Te_3$, $Bi_2Se_3$, BST, and BTSe have similar structures, we expect that the doping nature will be maintained in these binaries and ternaries similar to the dopant in $Bi_2Te_3$. The defect $E_{FORM}$ results are shown in Figure 8 for Bi- and Te-rich conditions. For F point impurity, $F_{Te(1)}$, $F_{Te(2)}$, and $F_{INT}$ are stable. For Cl, Br, and I, the substitutional defects at Te(2) are the most stable. From the density of states (DOS) analysis, we confirm the electrical properties of defects. In Figure 9, we show the total DOSs for $Bi_{96}Te_{144}$ without an impurity defect, $Bi_{96}Te_{143}Cl$ with single Cl substitutional defect at Te(2) site, $Bi_{96}Te_{144}Cl_{INT}$ with single interstitial defect at octahedral site. The Fermi level lies at the middle of the band gap for non-doped $Bi_2Te_3$. When Cl defect is generated, $Cl_{Te(2)}$ acts as a shallow donor with an excess electron at the conduction band minimum, while $Cl_{INT}$ acts as a shallow acceptor with an excess hole at the valence band maximum. The doping behavior is same for all halogen atoms we considered, F, Cl, Br, and I. All $M_{Te(2)}$ and $M_{INT}$ are shallow donors and acceptors, respectively. From the formation energy and DOS calculations results, we conclude that F is a bipolar defect. It can be a donor and acceptor and thereby it can be self-conpensated. However, Cl, Br, and I atoms are good shallow donors, preferring donor defects. For the high n-type carrier concentration in BTSe alloys, we recommend to use Cl, Br, or I dopants.

## 4. Conclusion

In conclusion, we have investigated the thermoelectric power factors of Bi-Sb-Te and Bi-Te-Se ternary alloys by performing the density functional theory calculations combined with Boltzmann transport equations. We show that the thermoelectric performances of Bi-Sb-Te and Bi-Te-Se alloys are comparable to the $Bi_2Te_3$ binary. The thermoelectric performance of p-type Bi-Sb-Te is superior to that of n-type Bi-Te-Se due to the longer electron relaxation time for BST and the poor in-plane transport properties originating from $Bi_2Se_3$. The electron transport anisotropy is severe for BTSe compared to BST. The optimal doping concentration for p-type Bi-Sb-Te is about $4\times10^{19}$ cm$^{-3}$, which is achievable by alloying. The optimal doping concentration for n-type Bi-Te-Se is about $6\times10^{19}$ cm$^{-3}$ or higher, which needs additional extrinsic dopant. The point defect formation energy calculations reveal that Cl, Br, and I impurities in $Bi_2Te_3$ and BTSe are potential candidate for n-type carrier source, while F is the self-compensating defect.

**Acknowledgements**

This work was supported by the Korea Institute of Energy Technology Evaluation and Planning (KETEP) and the Ministry of Trade, Industry & Energy (MOTIE) of the Republic of Korea (No. 20162000000910). B. Ryu and J. Chung waere also supported by This research was supported by Korea Electrotechnology Research Institute (KERI) Primary research program through the National Research Council of Science & Technology (NST) funded by the Ministry of Science, ICT and Future Planning(MSIP) (No. 17-12-N0101-38).

**References**

[1] D.M. Rowe, Thermoelectrics Handbook, CRC Press, 2006.

[2] G.J. Snyder, E.S. Toberer, Nat. Mater. 7 (2008) 105.

[3] H.J. Goldsmid, Introduction to Thermoelectricity, Springer, 2010, pp. 63-78.

[4] J. Garg, N. Bonini, B. Kozinsky, N. Marzari, Phys. Rev. Lett. 106 (2011) 045901.

[5] K.F. Hsu, S. Loo, F. Guo, W. Chen, J.S. Dyck, C. Uher, T. Hogan, E.K. Polychroniadis, M.G. Kanatzidis, Science 303 (2004) 818.

[6] P.F.P. Poudeu, J. D'Angelo, H. Kong, A. Downey, J. L. Short, R. Pcionek, T.P. Hogan, C. Uher, M.G. Kanatzidis, J. Am. Chem. Soc. 128 (2006) 14347.

[7] Y. Pei, J. Lensch-Falk, E.S. Toberer, D.L. Medlin, G.J. Snyder, Adv. Func. Mater. 21 (2011) 241.

[8] S.-H. Lo, J. He, K. Biswas, M.G. Kanatizidis, V.P. Dravid, Adv. Func. Mater. 22 (2012) 5175.

[9] Z. Tian, J. Gar, K. Esfarjani, T. Shiga, J. Shiomi, G. Chen, Phys. Rev. B 85 (2012) 184303.

[11] L.-D. Zhao, S. Hao, S.-H. Lo, C.-I. Wu, X. Zhou, Y. Lee, H. Li, K. Biswas, T.P. Hogan, C. Uher, C. Wolverton, V. P. Dravid, M.G. Kanatizidis, J. Am. Chem. Soc. 135 (2013) 7364.

[12] J. P. Heremans, V. Jovovic, E. S. Toberer, A. Saramat, K. Kurosaki, A. Charoenphakdee, S. Yamanaka, G.J. Snyder, Science 321 (2008) 554.

[13] Y. Pei, X. Shi, A. LaLonde, H. Wang, L. Chen, G. J. Snyder, Nature 473 (2011) 66.

[14] W. Liu, X. Tan, K. Yin, H. Liu, X. Tang, J. Shi, Q. Zhang, C. Uher, Phys. Rev. Lett. 108 (2012) 166601.

[15] S.J. Youn, A.J. Freeman, Phys. Rev. B 63 (2001) 085112

[16] T.J. Scheidemantel, C. Ambrosch-Draxl, T. Thonhauser, J.V. Badding, J.O. Sofo, Phys. Rev. B 68 (2003) 125210.

[17] G. Wang, T. Cagin, Phys. Rev. B 76 (2007) 075201.

[18] E. Kioupakis, M. L. Tiago, S.G. Louie, Phys. Rev. B 82 (2010) 245203.

[19] B. Ryu, B.-S. Kim, J. E. Lee, S.-J. Joo, B.-K. Min, H. W. Lee, S. D. Park, J. Kor. Phys. Soc. 68 (2016) 115.

[20] S. Park, B. Ryu, J. Kor. Phys. Soc. 69 (2016) 1683.

[21] B. Ryu, M.-W. Oh, J. Kor. Cer. Soc. 53 (2016) 273.

[22] J.K. Lee, S.D. Park, B.S. Kim. M.W. Oh, S.H. Cho, B.K. Min, H.W. Lee, M.H. Kim, Electron. Mater. Lett. 6 (2010) 201

[23] J.H. Son, M.W. Oh, B.S. Kim, S.D. Park, B.K. Min, M.H. Kim, H.W. Lee, J. Alloys Comp. 566 (2013) 168.

[24] S.I. Kim, K.H. Lee, H.A. Mun, H.S. Kim, S.W. Hwang, J.W. Roh, D.J. Yang, W.H. Shin, X.H. Li, Y.H. Lee, G.J. Snyder, S.W. Kim, Science 348 (2015) 109.

[25] M.W. Oh, J.H. Son, B.S. Kim, S.D. Park, B.K. Min, H.W. Lee, J. Appl. Phys. 115 (2014) 133706.

[26] M.S. Park, J.H. Song, J.E. Medvedeva, M. Kim, I.G. Kim, A.J. Freeman, Phys. Rev. B 81 (2010) 155211.

[27] N.F. Hinsche, B.Y. Yavorsky, I. Mertig, P. Zahn, Phys. Rev. B 84 (2011) 165214.

[28] X. Luo, M. B. Sullivan, S.Y. Quek, Phys. Rev. B 86 (2012) 184111.

[29] L.-P. Hu, T.-J. Zhu, Y.-G. Wang, H.-H. Xie, Z.-J. Xu, X.-B. Zhao, NPG Asia Mater. 6 (2014) e88.

[30] H.Y. Lv, H.J. Liu, L. Pan, Y.W. Wen, X.J. Tan, J. Shi, X.F. Tang, Appl. Phys. Lett. 96 (2010) 142101.

[31] M.W. Oh, B. Ryu, J.E. Lee, S.J. Joo, B.S. Kim, S.D. Park, B.K. Min, H.W. Lee, J. Nanoelectron. Optoelectron. 10 (2015) 391.

[32] P. Hohenberg, W. Kohn, Phys. Rev. 136 (1964) B864.

[33] W. Kohn, L.J. Sham, Phys. Rev. 140 (1965) A1133.

[34] P. E. Blöchl, Phys. Rev. B 50 (1994) 17953.

[35] J. P. Perdew, K. Burke, M. Ernzerhof, Phys. Rev. Lett. 77 (1996) 3865.

[36] G. Kresse, J. Furthmüller, Phys. Rev. B 54 (1996) 11169.

[37] G. Kresse, D. Joubert, Phys. Rev. B 59 (1999) 1758.

[38] G.K.H. Madsen, D.J. Singh, Comput. Phys. Commun. 175 (2006) 67.

[39] Interdisciplinary Centre for Advanced Materials Simulation, BOLTZTRAP program download, http://www.icams.de/content/research/software-development/boltztrap/, 2017 (accessed 17.04.05).

[40] J.-H. Bahk, A. Shakouri, Phys. Rev. B 93 (2016) 165209.

[41] H.J. Goldsmid, Introduction to Thermoelectricity, Springer, 2010, pp. 51-53.

[42] I.G. Austin, Proc. Phys. Soc. London 72 (1958) 545.

[43] H. Landolt, R. Börnstein, Landolt-Börnstein - Group III Condensed Matter, 41C, 1998, pp. 1-6.

[44] Y.S. Hor, A. Richardella, P. Roushan, Y. Xia, J.G. Checkelsky, A. Yazdani, M.Z. Hasan, N.P. Ong, R.J. Cava, Phys. Rev. B 79 (2009) 195208.

[45] D. Parker, D.J. Singh, Phys. Rev. X 1 (2011) 021005.

[46] O.V. Yzayev, E. Kioupakis, J.E. Moore, S.G. Louie, Phys. Rev. B 85 (2012) 161101(R).

[47] B. Poudel, Q. Hao, Y. Ma, Y. Lan, A. Minnich, B. Yu, X. Yan, D. Wang, A. Muto, D. Vashaee, X. Chen, J. Liu, M.S. Dresselhaus, G. Chen, Z. Ren, Science 320 (2008) 634.

[48] S. Hwang, S.I. Kim, K. Ahn, J.W. Roh, D.J. Yang, S. M. Lee, K.H. Lee, J. Electron. Mater. 42 (2013) 1411.

[49] K.H. Lee, S.-M. Choi, J.W. Roh, S. Hwang, S.I. Kim, W.H. Shin, H.J. Park, J.H Lee, S.W. Kim, D.J. yang, J. Electron. Mater. 44 (2015) 1531.

[50] J.K. Lee, personal communication.

[51] G.-E. Lee, Synthesis and Thermoelectric Properties of $Bi_2Te_3$-$Bi_2Se_3$ Solid Solutions, Thesis, Korea National University of Transportation, 25015.

[52] J.H. Son, personal communication.

[53] G.-E. Lee, personal communication.]

[54] S.B. Zhang, J.E. Northrup, Phys. Rev. Lett. 67 (1991) 2339.

[55] B. Ryu, M.-W. Oh, J.K. Lee, J.E. Lee, S.-J. Joo, B.-S. Kim, B.-K. Min, H.-W. Lee, S.D. Park, J. Appl. Phys. 118 (2015) 015705.

[56] B.Y. Yavorsky, N.F. Hinsche, I. Mertig, P. Zhan, Phys. Rev. B 84 (2011) 165208.

[57] W.-S. Liu, Q. Zhang, Y. Lan, S. Chen, X. Yan, Q. Zhang, H. Wang, D. Wang, G. Chen, Z. Ren, Adv. Energy Mater. 1 (2011) 577.

[58] G.-E. Lee, I-.H. Kim, Y. S. Lim, W.-S. Seo, B.-J. Choi, C.-W. Hwang, J. Kor. Phsy. Soc. 64 (2014) 1692.

[59] G.-E. Lee, I.-H. Kim, Y. S. Lim, W.-S. Seo, B.-J. Choi, C.-W. Hwang, J. Kor. Phys. Soc. 65 (2014) 696.

[60] K.H. Lee, S.I. Kim, H. Mun, B. Ryu, S.-M. Choi, H.J. Park, S. Hwang, S.W. Kim, J. Mater. Chem. C 3 (2015) 10604.

**Tables**

Table 1 The details of BST atomic models are shown, including lattice parameters, internal coordinates, and the number of atoms in the given supercell.

| Model Name | x of $(Bi_{1-x}Sb_x)_2Te_3$ | Structure Parameter | | | | Number of atoms in supercell | | | Number of configuration |
|---|---|---|---|---|---|---|---|---|---|
| | | a (Å) | c (Å) | u | v | Bi | Sb | Te | |
| P0 | 0.000 | 4.3835 | 30.4870 | 0.4000 | 0.2096 | 8 | 0 | 12 | 1 |
| P1 | 0.125 | 4.3668 | 30.4761 | 0.3998 | 0.2098 | 7 | 1 | 12 | 1 |
| P2 | 0.250 | 4.3221 | 30.0252 | 0.4002 | 0.2099 | 6 | 2 | 12 | 1 |
| P3-1, P3-2 | 0.375 | 4.2914 | 29.7943 | 0.4003 | 0.2101 | 5 | 3 | 12 | 2 |
| P4-1, P4-2 | 0.500 | 4.2607 | 29.5634 | 0.4003 | 0.2103 | 4 | 4 | 12 | 2 |
| P5-1, P5-2 | 0.625 | 4.2300 | 29.3325 | 0.4004 | 0.2104 | 3 | 5 | 12 | 2 |
| P6 | 0.750 | 4.1994 | 29.1016 | 0.4005 | 0.2106 | 2 | 6 | 12 | 1 |
| P7 | 0.875 | 4.1687 | 28.8707 | 0.4006 | 0.2107 | 1 | 7 | 12 | 1 |
| P8 | 1.000 | 4.2500 | 30.4000 | 0.3987 | 0.2110 | 0 | 8 | 12 | 1 |

Table 2 The details of BTSe atomic models are shown, including lattice parameters, internal coordinates, and the number of atoms in the given supercell.

| Model Name | y of Bi$_2$(Te$_{1-y}$Se$_y$)$_3$ | Structure Parameters | | | | Number of atoms in supercell | | | | |
|---|---|---|---|---|---|---|---|---|---|---|
| | | a | c | u | v | Bi | Te | Se | Se(1) | Se(2) |
| N0 | 0.0000 | 4.3835 | 30.4870 | 0.4000 | 0.2096 | 8 | 12 | 0 | 0 | 0 |
| N1-1 | 0.0833 | 4.3630 | 30.3331 | 0.4001 | 0.2097 | 8 | 11 | 1 | 1 | 0 |
| N1-2 | | | | | | | | | 0 | 1 |
| N2-1 | 0.1667 | 4.3426 | 30.1791 | 0.4001 | 0.2098 | 8 | 10 | 2 | 1 | 1 |
| N2-2 | | | | | | | | | 0 | 2 |
| N3-1 | 0.2500 | 4.3221 | 30.0252 | 0.4002 | 0.2099 | 8 | 9 | 3 | 2 | 1 |
| N3-2 | | | | | | | | | 0 | 3 |
| N4-1 | 0.3333 | 4.3017 | 29.8713 | 0.4002 | 0.2100 | 8 | 8 | 4 | 2 | 2 |
| N4-2 | | | | | | | | | 3 | 1 |
| N4-3 | | | | | | | | | 0 | 4 |
| N5 | 0.4167 | 4.2812 | 29.7173 | 0.4003 | 0.2102 | 8 | 7 | 5 | 4 | 1 |
| N6-1 | 0.5000 | 4.2607 | 29.5634 | 0.4003 | 0.2103 | 8 | 6 | 6 | 4 | 2 |
| N6-2 | | | | | | | | | 4 | 2 |
| N6-3 | | | | | | | | | 2 | 4 |
| N7-1 | 0.5833 | 4.2403 | 29.4095 | 0.4004 | 0.2104 | 8 | 5 | 7 | 4 | 3 |
| N7-2 | | | | | | | | | 3 | 4 |
| N8-1 | 0.6667 | 4.2198 | 29.2555 | 0.4004 | 0.2105 | 8 | 4 | 8 | 5 | 3 |
| N8-2 | | | | | | | | | 4 | 4 |
| N9-1 | 0.7500 | 4.1994 | 29.1016 | 0.4005 | 0.2106 | 8 | 3 | 9 | 6 | 3 |
| N9-2 | | | | | | | | | 5 | 4 |
| N10-1 | 0.8333 | 4.1789 | 28.9477 | 0.4006 | 0.2107 | 8 | 2 | 10 | 7 | 3 |
| N10-2 | | | | | | | | | 6 | 4 |
| N11-1 | 0.9167 | 4.1584 | 28.7937 | 0.4006 | 0.2108 | 8 | 1 | 11 | 7 | 4 |
| N11-2 | | | | | | | | | 8 | 3 |
| N12 | 1.0000 | 4.1380 | 28.6398 | 0.4007 | 0.2109 | 8 | 0 | 12 | 8 | 4 |

**Table 3** Optimal carrier concentration to maximize power factor at 300 K are calculated for all ternary BST alloys at 300K. Here we search the maximum power factor for hole carrier concentration less than 2 x $10^{20}$ cm$^{-3}$ and then find the corresponding optimal hole carrier concentration.

| Model Name | x of $(Bi_{1-x}Sb_x)_2Te_3$ | Optimal hole concentration | | |
|---|---|---|---|---|
| | | $PF_{inp}^{MAX}$ | $PF_{outp}^{MAX}$ | $<PF>^{MAX}$ |
| P0 | 0.000 | 3.9E+19 | 4.6E+19 | 3.9E+19 |
| P1 | 0.125 | 3.7E+19 | 4.3E+19 | 3.7E+19 |
| P2 | 0.250 | 3.5E+19 | 4.1E+19 | 3.5E+19 |
| P3-1 | 0.375 | 3.5E+19 | 4.1E+19 | 3.5E+19 |
| P3-2 | | 3.4E+19 | 4.0E+19 | 3.4E+19 |
| P4-1 | 0.500 | 3.3E+19 | 3.3E+19 | 3.3E+19 |
| P4-2 | | 3.2E+19 | 3.2E+19 | 3.2E+19 |
| P5-1 | 0.625 | 3.1E+19 | 3.1E+19 | 3.1E+19 |
| P5-2 | | 2.9E+19 | 3.4E+19 | 2.9E+19 |
| P6 | 0.750 | 2.9E+19 | 2.9E+19 | 2.9E+19 |
| P7 | 0.875 | 2.8E+19 | 3.3E+19 | 2.8E+19 |
| P8 | 1.000 | 3.2E+19 | 3.2E+19 | 3.2E+19 |

Table 4 Optimal carrier concentration to maximize power factor at 300 K are calculated for all ternary BTSe alloys at 300K. Here we search the maximum power factor for hole carrier concentration less than 2 x $10^{20}$ cm$^{-3}$ and then find the corresponding optimal hole carrier concentration.

| Model Name | y of $Bi_2(Te_{1-y}Se_y)_3$ | Optimal electron concentration | | |
|---|---|---|---|---|
| | | $PF_{inp}^{MAX}$ | $PF_{outp}^{MAX}$ | $<PF>^{MAX}$ |
| N0 | 0.000 | 4.7E+19 | >2E+20 | 4.0E+19 |
| N1-1 | 0.083 | 3.9E+19 | >2E+20 | 3.3E+19 |
| N1-2 | | 6.2E+19 | >2E+20 | 7.1E+19 |
| N2-1 | 0.167 | 4.3E+19 | >2E+20 | 4.3E+19 |
| N2-2 | | 9.8E+19 | >2E+20 | 1.4E+20 |
| N3-1 | 0.250 | 3.9E+19 | >2E+20 | 3.3E+19 |
| N3-2 | | 1.3E+20 | 1.7E+20 | 1.6E+20 |
| N4-1 | 0.333 | 4.6E+19 | 2.6E+19 | 4.6E+19 |
| N4-2 | | 4.0E+19 | 2.8E+19 | 3.3E+19 |
| N4-3 | | 1.5E+20 | 1.5E+20 | 1.5E+20 |
| N5 | 0.417 | 3.9E+19 | 3.9E+19 | 3.9E+19 |
| N6-1 | 0.500 | 4.0E+19 | 4.0E+19 | 4.0E+19 |
| N6-2 | | 4.4E+19 | 3.7E+19 | 3.7E+19 |
| N6-3 | | 6.3E+19 | >2E+20 | 6.3E+19 |
| N7-1 | 0.583 | 4.8E+19 | 4.1E+19 | 4.8E+19 |
| N7-2 | | 4.9E+19 | 4.1E+19 | 4.9E+19 |
| N8-1 | 0.667 | 5.5E+19 | 6.1E+19 | 5.5E+19 |
| N8-2 | | 4.8E+19 | 4.0E+19 | 4.8E+19 |
| N9-1 | 0.750 | 4.5E+19 | 4.5E+19 | 4.5E+19 |
| N9-2 | | 5.8E+19 | 5.1E+19 | 5.1E+19 |
| N10-1 | 0.833 | 7.2E+19 | 5.5E+19 | 7.2E+19 |
| N10-2 | | 5.4E+19 | 4.2E+19 | 5.4E+19 |
| N11-1 | 0.917 | 7.3E+19 | 2.8E+19 | 6.1E+19 |
| N11-2 | | 9.8E+19 | 1.5E+19 | 9.8E+19 |
| N12 | 1.000 | 5.5E+19 | 1.6E+19 | 4.6E+19 |

**Figure Captions**

Figure 1 The directional average of Seebeck coefficient as a function of carrier concentration is shown for Bi2Te3. In (a), the Seebek coefficient is calculated with band gaps from the DFT-PBE calculation (0.10eV) and the experiment value (0.16 eV). In (b), The Seebeck coefficient function is calculated for various temperatures, 300, 400, and 500 K, with the experimental band gap.

Figure 2 The directional average of Seebeck coefficient at 300 K is calculated and drawn (a) for p-type and (b) for n-type binary phases of $Bi_2Te_3$, $Sb_2Te_3$, $Bi_2Se_3$, and $Sb_2Se_3$ in tetradymite phase.

Figure 3 The mixing Free energy G of (a) BST and (b) BTSe solutions. The symbols represent the G of models considered here. The solid, dot, and sashed lines represent the fitted G curve.

Figure 4 (a) The PFs along in-plane and (b) along out-of-plane directions are calculated for BST.

Figure 5 (a) Maximum PF when **n** is less than $2\times10^{20}$ cm$^{-3}$ for BST. (b) Maximum PF when **n** is less than $2\times10^{20}$ cm$^{-3}$ for BST. $PF_{INP}^{MAX}$, $PF_{OUTP}^{MAX}$, and $<PF>^{MAX}$ are maxima of $PF_{INP}$, maximum of out-of-plane $PF_{OUTP}$, the maximum of directional averaged PF, $<PF>$. $<PF>$ is defined as $<PF> = 1/3 \times ( 2\ PF_{INP} + 1\ PF_{OUTP})$.

Figure 6 Temperature dependent $PF_{inp}$ for (a) p-$Bi_2Te_3$ and (b) p-$(Bi_{0.25}Sb_{0.75})_2Te_3$, and $PF_{outp}$ for (c) p-$Bi_2Te_3$ and (d) p-$(Bi_{0.25}Sb_{0.75})_2Te_3$, when **n** = 1, 2, 4, 6, $8\times10^{19}$ cm$^{-3}$.

Figure 7 Temperature dependent $PF_{inp}$ for (a) n-$Bi_2Te_3$ and (b) n-$Bi_2(Te_{0.917}Se_{0.083})_3$, and $PF_{outp}$ for (c) n-$Bi_2Te_3$ and (d) n-$Bi_2(Te_{0.917}Se_{0.083})_3$, when **n** = 1, 2, 4, 6, $8\times10^{19}$ cm$^{-3}$.

Figure 8 Defect formation energies are calculated for various point impurities in $Bi_2Te_3$: M substitutional at Bi ($M_{Bi}$), M substitutional at Te(1) ($M_{Te(1)}$), F substitutional at Te(2) ($M_{Te(2)}$), and F interstitial ($M_{INT}$), where M = F, Cl, Br, I. We consider both the Bi and Te-rich conditions.

Figure 9 Total Density of States for (a) $Bi_{96}Te_{144}$, (b) $Bi_{96}Te_{143}Cl_{Te(1)}$ and (c) $Bi_{96}Te_{144}Cl_{INT}$ are shown by solid lines. The Fermi levels, denoted by vertical dotted lines, set to zero.

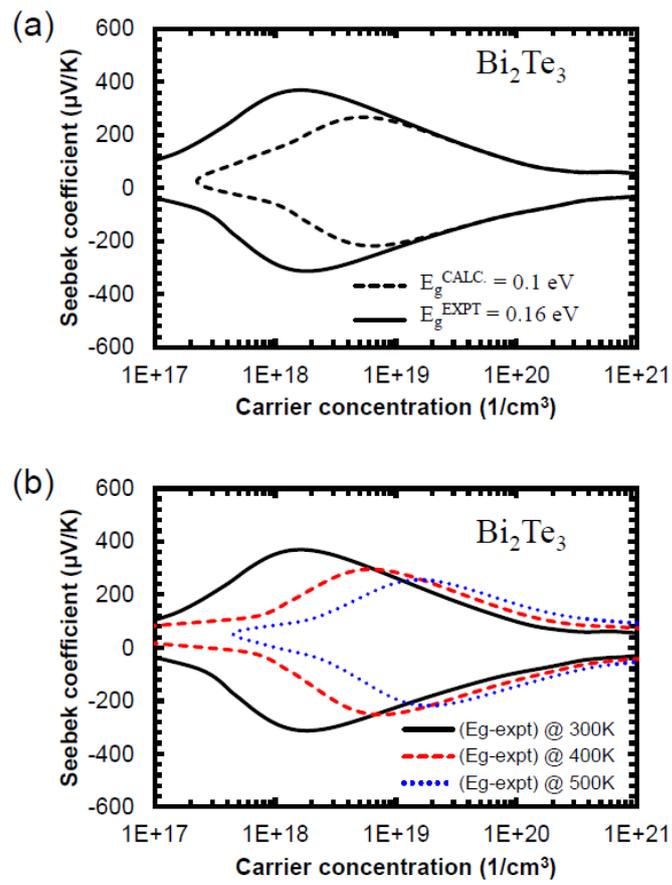

Figure 1

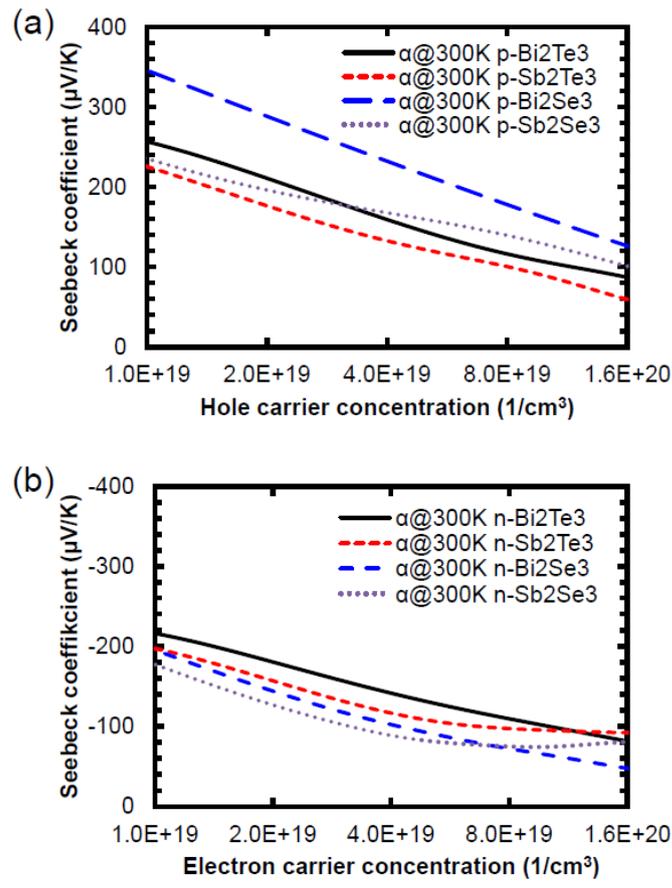

Figure 2

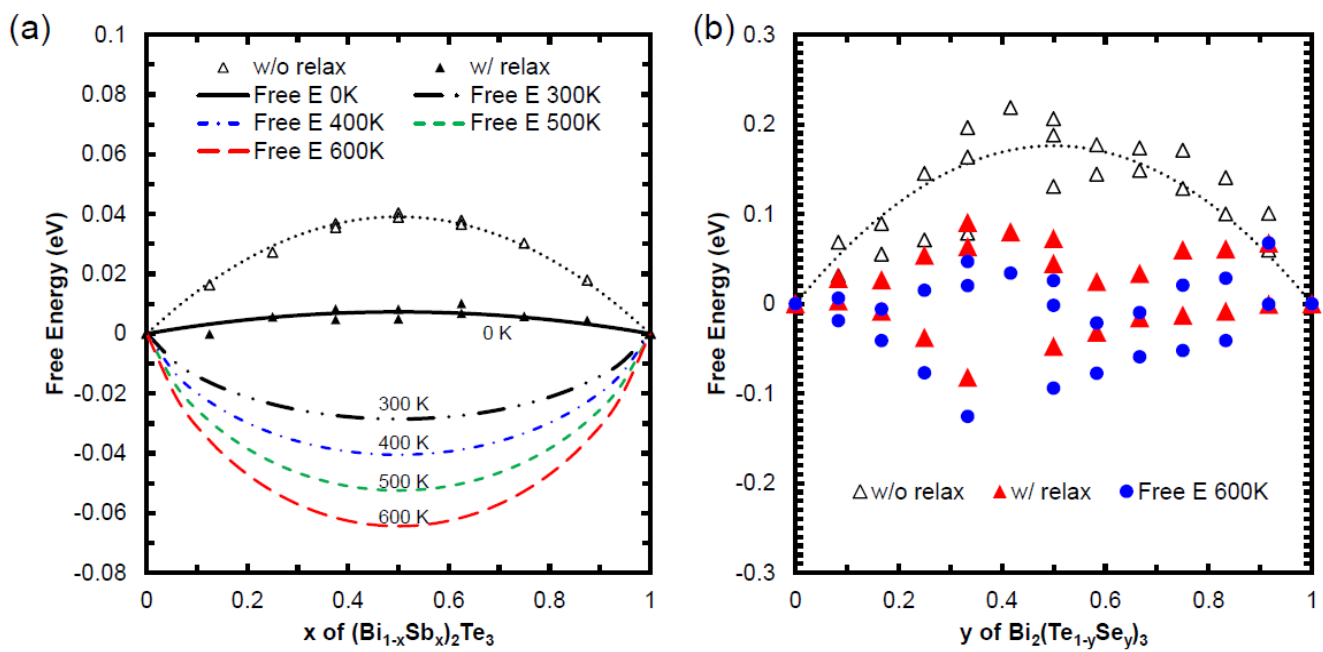

Figure 3

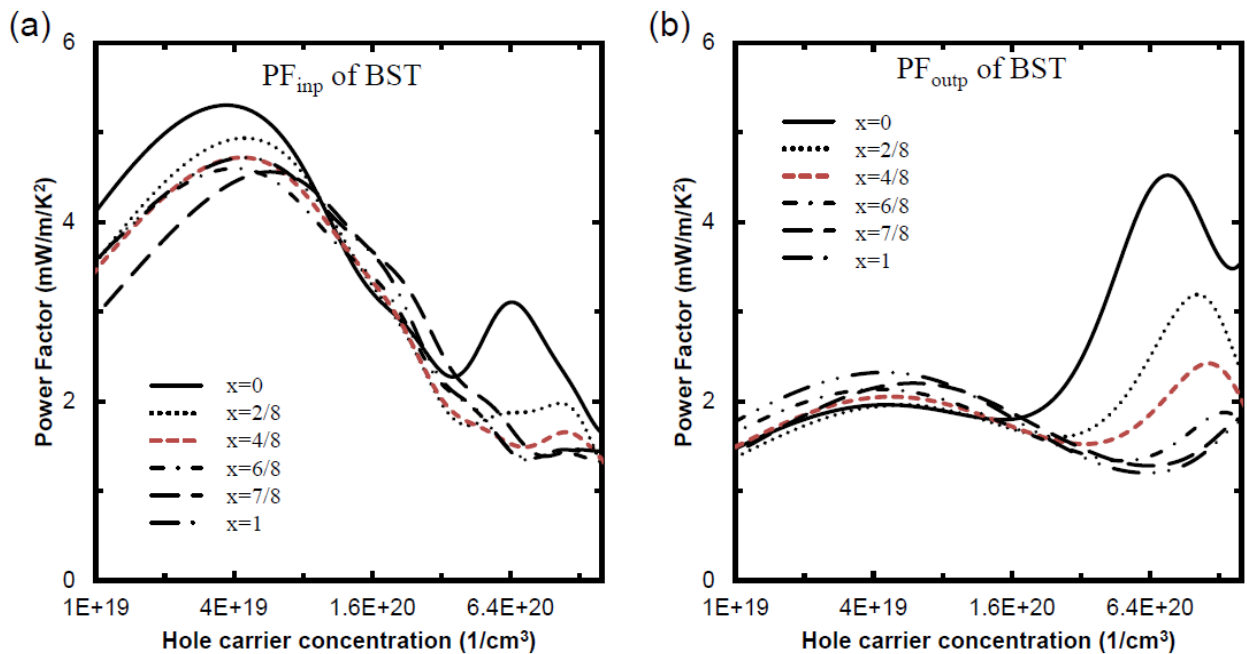

Figure 4

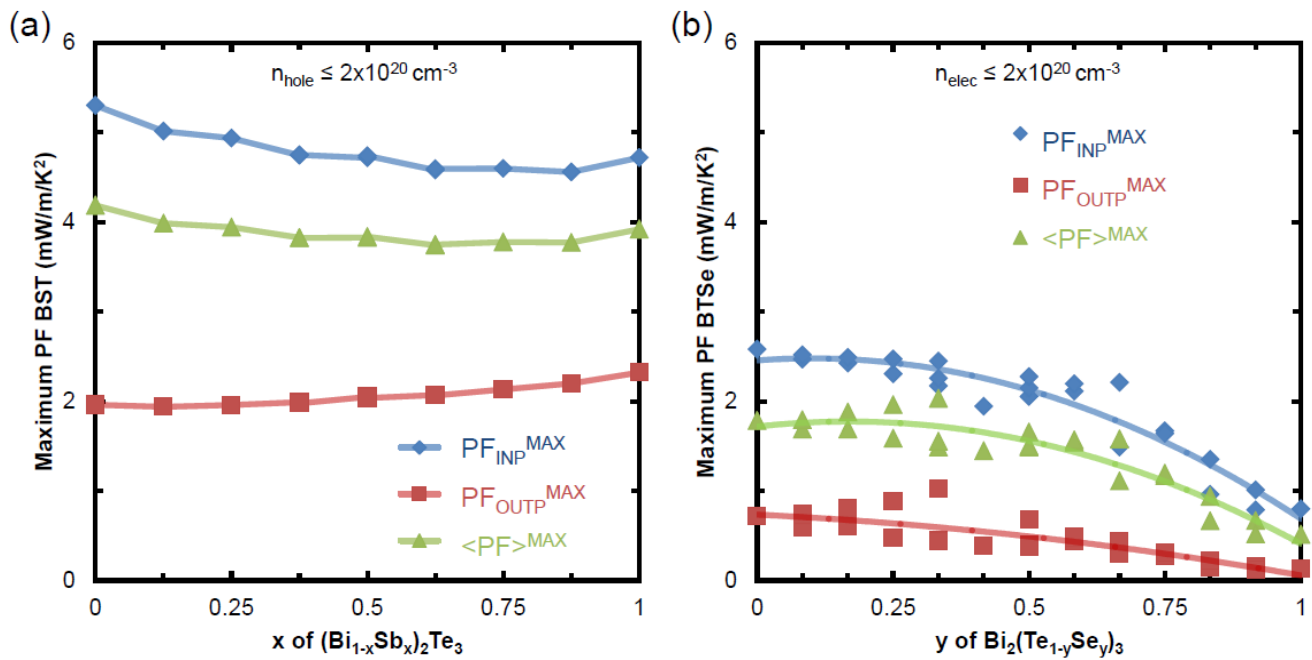

Figure 5

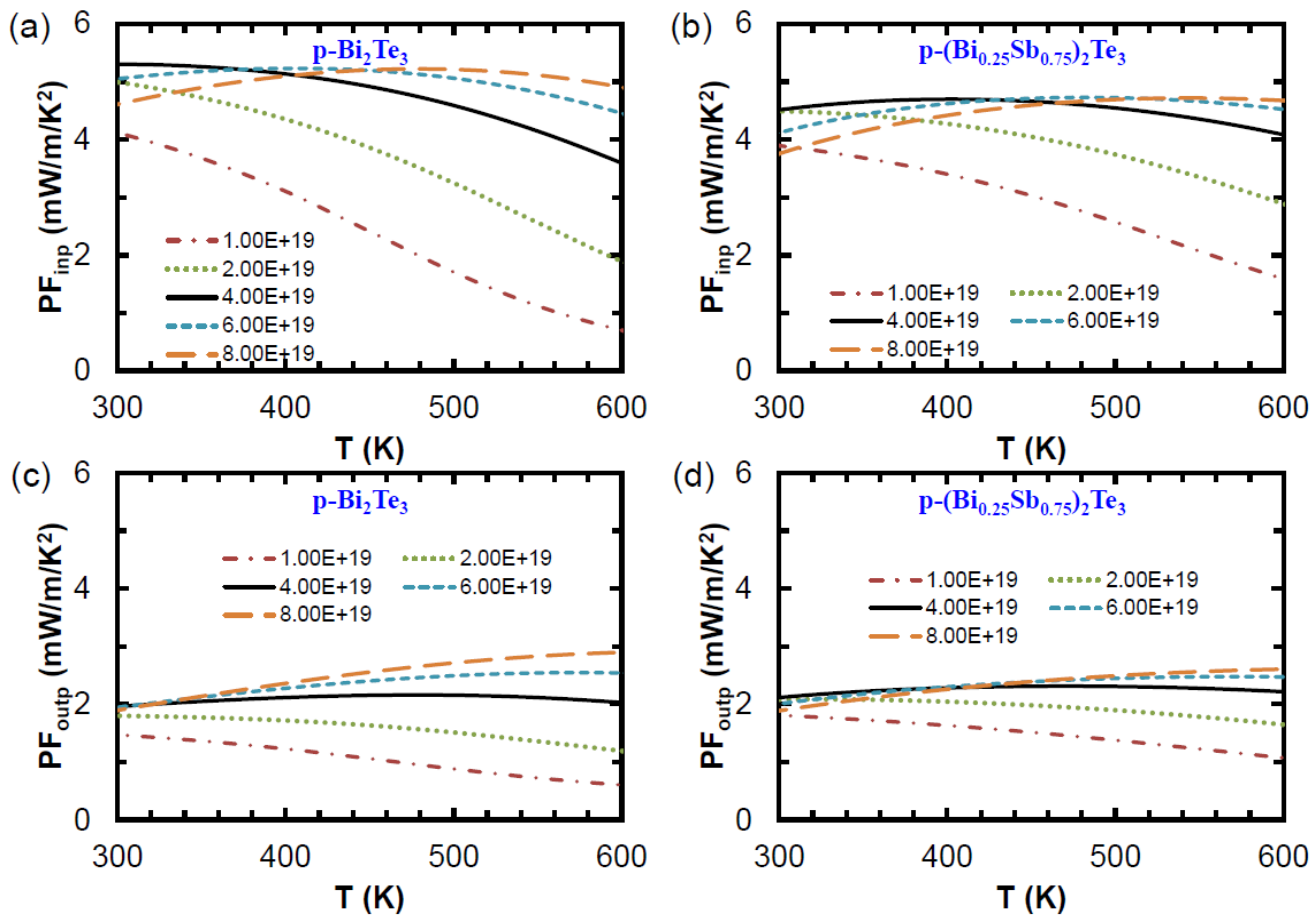

Figure 6

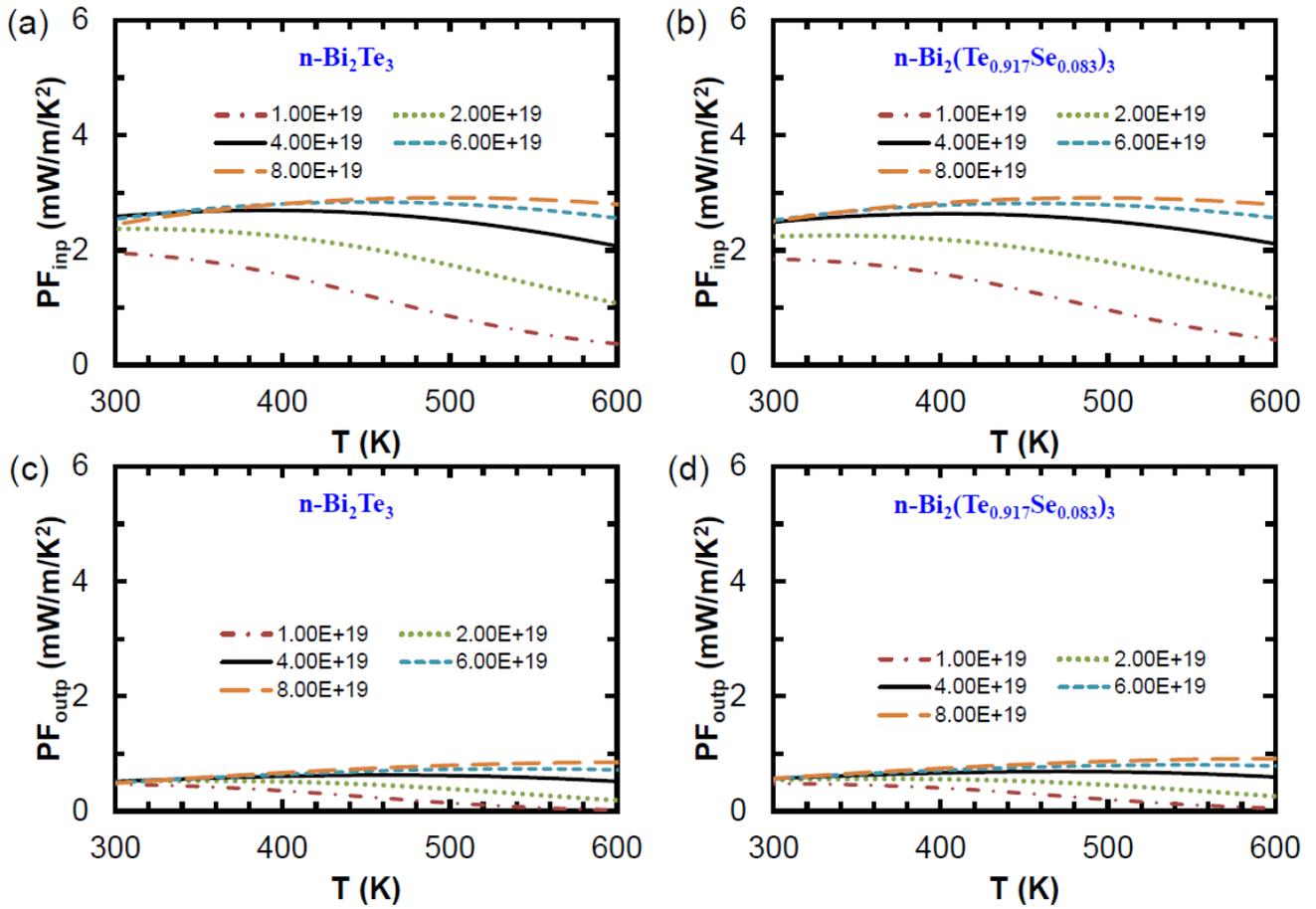

Figure 7

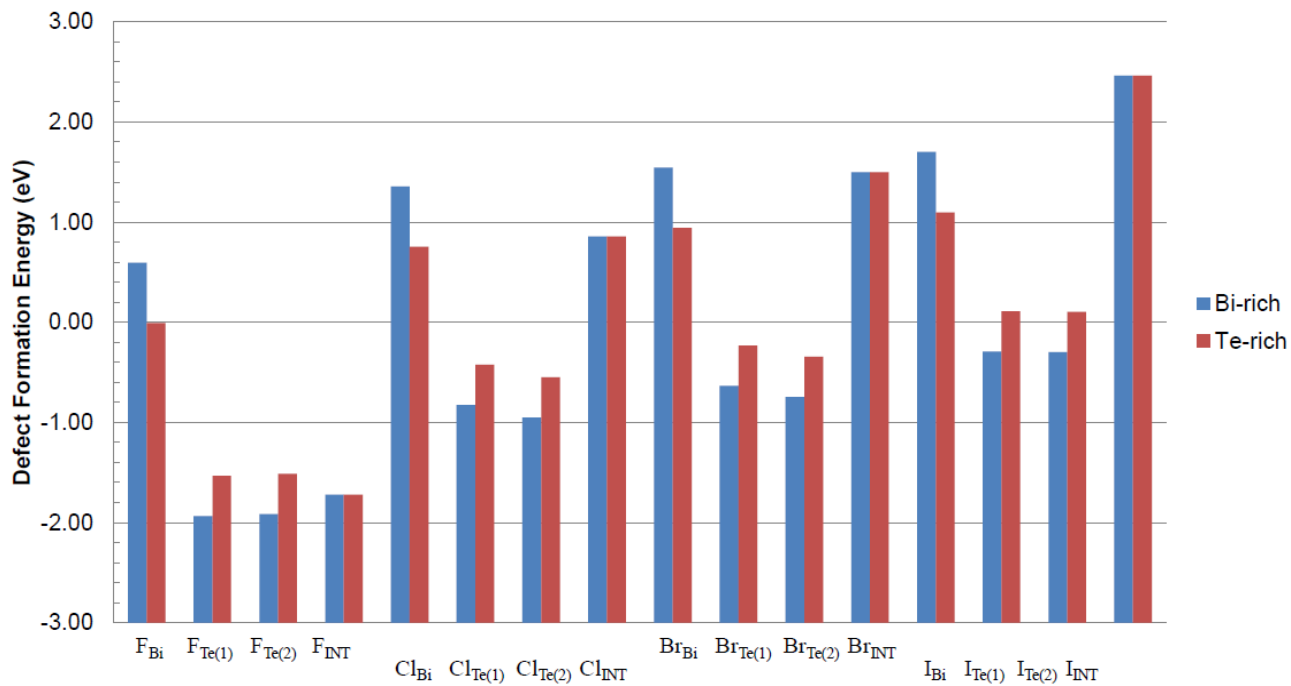

Figure 8

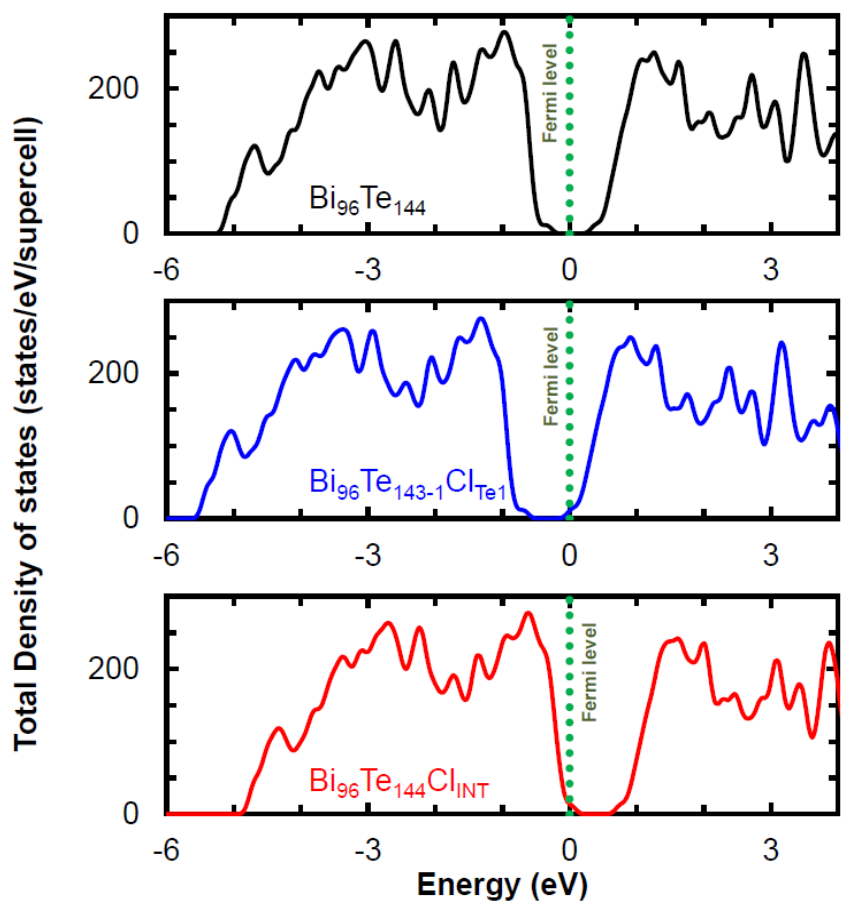

Figure 9